\documentclass[a4paper]{report}
\usepackage[utf8]{inputenc}
\usepackage[T1]{fontenc}
\usepackage{RJournal}
\usepackage{amsmath,amssymb,array}
\usepackage{booktabs}

\usepackage{subfig}

\begin{document}

\sectionhead{Contributed research article}
\volume{XX}
\volnumber{YY}
\year{20ZZ}
\month{AAAA}

\begin{article}
\title{Accessible Computation of Tight Symbolic Bounds on Causal Effects
using an Intuitive Graphical Interface}
\author{by Gustav Jonzon, Michael C Sachs, and Erin E Gabriel}

\maketitle

\abstract{%
Strong untestable assumptions are almost universal in causal point
estimation. In particular settings, bounds can be derived to narrow the
possible range of a causal effect. Symbolic bounds apply to all settings
that can be depicted using the same directed acyclic graph (DAG) and for
the same effect of interest. Although the core of the methodology for
deriving symbolic bounds has been previously developed, the means of
implementation and computation have been lacking. Our \texttt{R}-package
\CRANpkg{causaloptim} \citep{causaloptim} aims to solve this usability
problem by implementing the method of \citep{generalcausalbounds} and
providing the user with a graphical interface through \CRANpkg{shiny}
that allows for input in a way that most researchers with an interest in
causal inference will be familiar; a DAG (via a point-and-click
experience) and specifying a causal effect of interest using familiar
counterfactual notation.
}

\hypertarget{introduction-and-background}{%
\section{Introduction and
Background}\label{introduction-and-background}}

A common goal in many different areas of scientific research is to
determine causal relationships between one or more exposure variables
and an outcome. Prior to any computation or inference, we must clearly
state all assumptions made, i.e., all subject matter knowledge
available, regarding the causal relationships between the involved
variables as well as any additional variables, called confounders, that
may not be measured but influence at least two other variables of
interest. These assumptions are usually encoded in a causal directed
acyclic graph (DAG), with directed edges encoding direct causal
influences, which conveniently depicts all relevant information and has
become a familiar tool in applied research \citep{greenland1999causal}.
Such a DAG not only clearly states the assumptions made by the
researcher, but also comes with a sound methodology for causal
inference, in the form of identification results as well as derivation
of causal estimators \citep{pearl2009causality}.

Unfortunately, point identification of a desired causal effect typically
requires an assumption of no unmeasured confounders, in some form. When
there are unmeasured confounders, it is sometimes still possible to
derive bounds on the effect, i.e., a range of possible values for the
causal effect in terms of the observed data distribution. Symbolic
bounds are algebraic expressions for the bounds on the causal effect
written in terms of probabilities that can be estimated using observed
data. Alexander Balke and Judea Pearl first used linear programming to
derive tight symbolic bounds in a simple binary instrumental variable
setting \citep{balke1997bounds}. Balke wrote a program in \texttt{C++}
to take a linear programming problem as text file input, perform
variable reduction, conversion of equality constraints into inequality
constraints, and perform the vertex enumeration algorithm of
\citep{mattheiss1973algorithm}. This program has been used by
researchers in the field of causal inference
\citep{balke1997bounds, cai2008bounds, sjolander2009bounds, sjolander2014bounds}
but it is not particularly accessible because of the technical challenge
of translating the DAG plus causal query into the constrained
optimization problem and to determine whether it is linear. Moreover,
the program is not optimized and hence does not scale well to more
complex problems. Since they only cover a simple instrumental variable
setting, it has also not been clear to what extent their techniques
extend to more general settings, nor how to apply them to more complex
queries. Thus, applications of this approach have been limited to a
small number of settings and few attempts to generalize the method to
more widely applicable settings have been made.

Recent developments have expanded the applicability by generalizing the
techniques and the causal DAGs and effects to which they apply
\citep{generalcausalbounds}. These new methods have been applied in
novel observational and experimental settings
\citep{gabriel2020causal, gabriel2021nonparametric, gabriel2022sharp}.
Moreover, through the \texttt{R} package \CRANpkg{causaloptim}
\citep{causaloptim}, these computations are now accessible. With
\CRANpkg{causaloptim}, the user needs only to give input in a way they
would usually express their causal assumptions and state their target
causal estimand; through a DAG and counterfactual expression. Providing
DAGs through textual input is an awkward experience for most users, as
DAGs are generally communicated pictorially. Our package
\CRANpkg{causaloptim} provides a user-friendly graphical interface
through a web browser, where the user can draw their DAG in a way that
is familiar to them. The methodology that underpins
\CRANpkg{causaloptim} is not universal however; some restrictions on the
DAG and query are imposed. These are validated and communicated to the
user through the graphical interface, which guides the user through
providing the DAG and query, adding any extra conditions beyond those
encoded in the DAG, computing, interpreting and exporting the bounds for
various further analyses.

There exist few other \texttt{R}-packages related to causal bounds and
none to our knowledge for computation of symbolic bounds.
\CRANpkg{bpbounds} \citep{bpbounds-package} provides a text-based
interface to compute numeric bounds for the original single instrumental
variable example of Balke and Pearl and extends this by being able to
compute bounds given different types of data input including a ternary
rather than binary instrument. There is also a standalone program
written in \texttt{Java} by the \texttt{TETRAD\ Project}
(\url{https://github.com/cmu-phil/tetrad}) that includes a graphical
user interface and has a wrapper for \texttt{R}. Its focus, however, is
on causal discovery in a given sample data set, and although it can also
compute bounds, it can do so only numerically for the given data set.

In this paper we describe our \texttt{R} package \CRANpkg{causaloptim},
first focusing on the graphical and programmatic user interfaces in the
next 2 sections. Then we highlight some of our interesting functions and
data structures that may be useful in other contexts. We provide a
summary of the theoretical background and methods, while referring to
the companion paper \citep{generalcausalbounds} for the details. We
illustrate the use of the package with some numeric examples and close
with a discussion and summary.

\hypertarget{graphical-user-interface}{%
\section{Graphical User Interface}\label{graphical-user-interface}}

In the following, we will work through the binary instrumental variable
example, where we have 3 observed binary variables \(X\), \(Y\), \(Z\),
and we want to determine the average causal effect of \(X\) on \(Y\)
given by total causal risk difference, in the presence of unmeasured
confounding by \(U_R\) and an instrumental variable \(Z\). Our causal
DAG is given by \(Z\to X\to Y\) and \(X\leftarrow U_R\to Y\) and our
causal query is \(P(Y(X=1)=1)-P(Y(X=0)=1)\), where we use \(Y(X = x)\)
to denote the potential outcome for \(Y\) if \(X\) were intervened upon
to have value \(x\).

\CRANpkg{causaloptim} includes a graphical user interface using
\CRANpkg{shiny} \citep{shiny}. The interface is launched in the user's
default web browser by calling \texttt{specify\_graph()}. Once the
\CRANpkg{shiny} app is launched, the user is presented with an
interactive display as shown in Figure \ref{fig:InterfaceStart}, in
which they can draw their causal DAG. This display is divided into a
left side \(\mathcal{L}\) and right side \(\mathcal{R}\) to classify the
vertices according to the class of DAGs that the method covers. In
particular, the existence of unmeasured confounders is assumed within
each of these sides, but not between them, and any causal influence
between the two sides must originate in \(\mathcal{L}\). Thus, for the
example, we would want to put the instrumental variable on the left
side, but the exposure and outcome on the right side. In the web version
of this article an interactive version of this interface is shown at the
end of this section.

\begin{Schunk}
\begin{figure}

{\centering \includegraphics[width=1\linewidth]{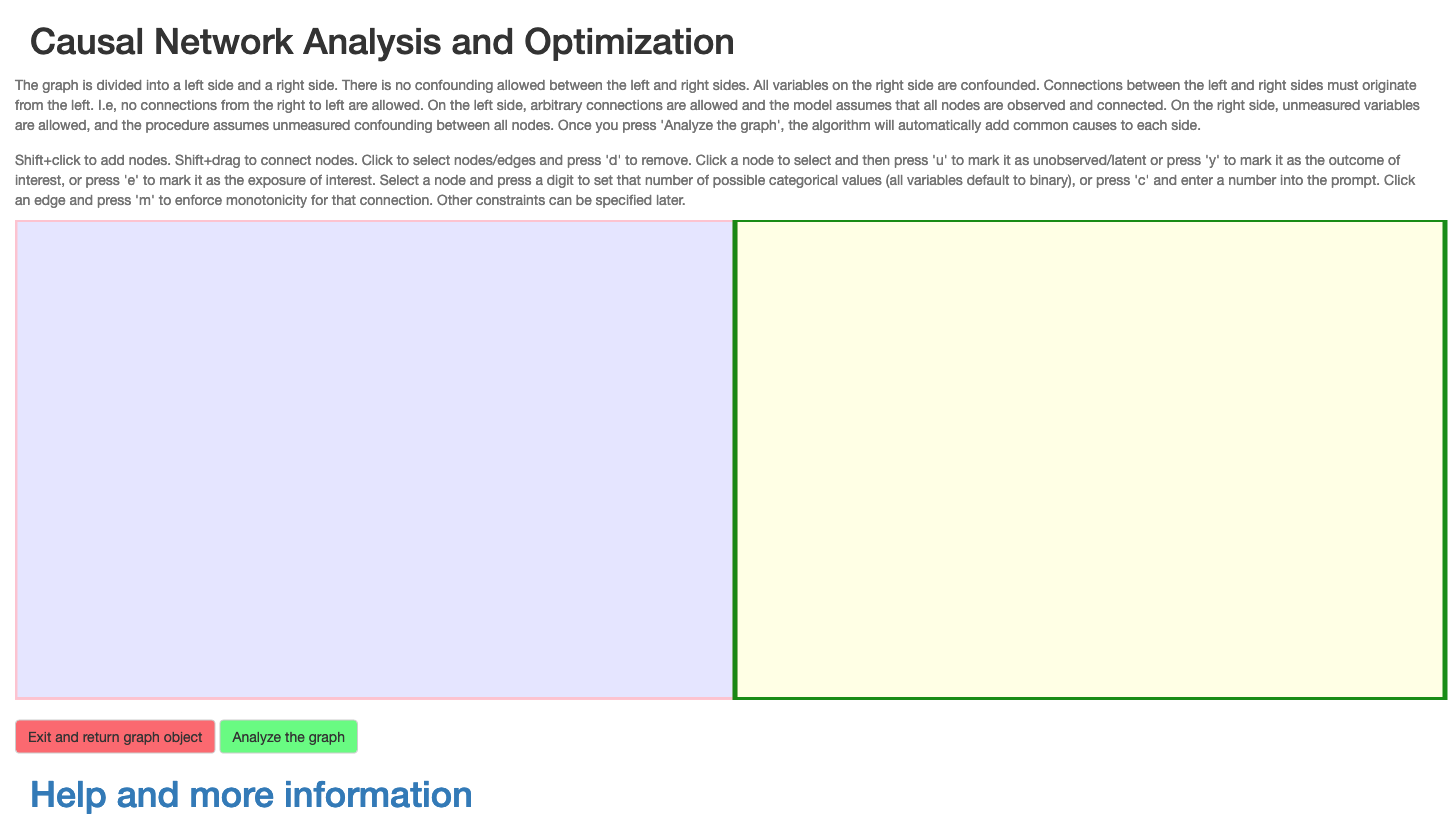} 

}

\caption[The Shiny web interface at lauch]{The Shiny web interface at lauch}\label{fig:InterfaceStart}
\end{figure}
\end{Schunk}

\hypertarget{specifying-the-setting-by-drawing-a-causal-diagram-and-adding-attributes}{%
\subsection{Specifying the setting by drawing a causal diagram and
adding
attributes}\label{specifying-the-setting-by-drawing-a-causal-diagram-and-adding-attributes}}

\begin{Schunk}
\begin{figure}
\subfloat[Adding and naming variables\label{fig:DAG-1}]{\includegraphics[width=0.5\linewidth]{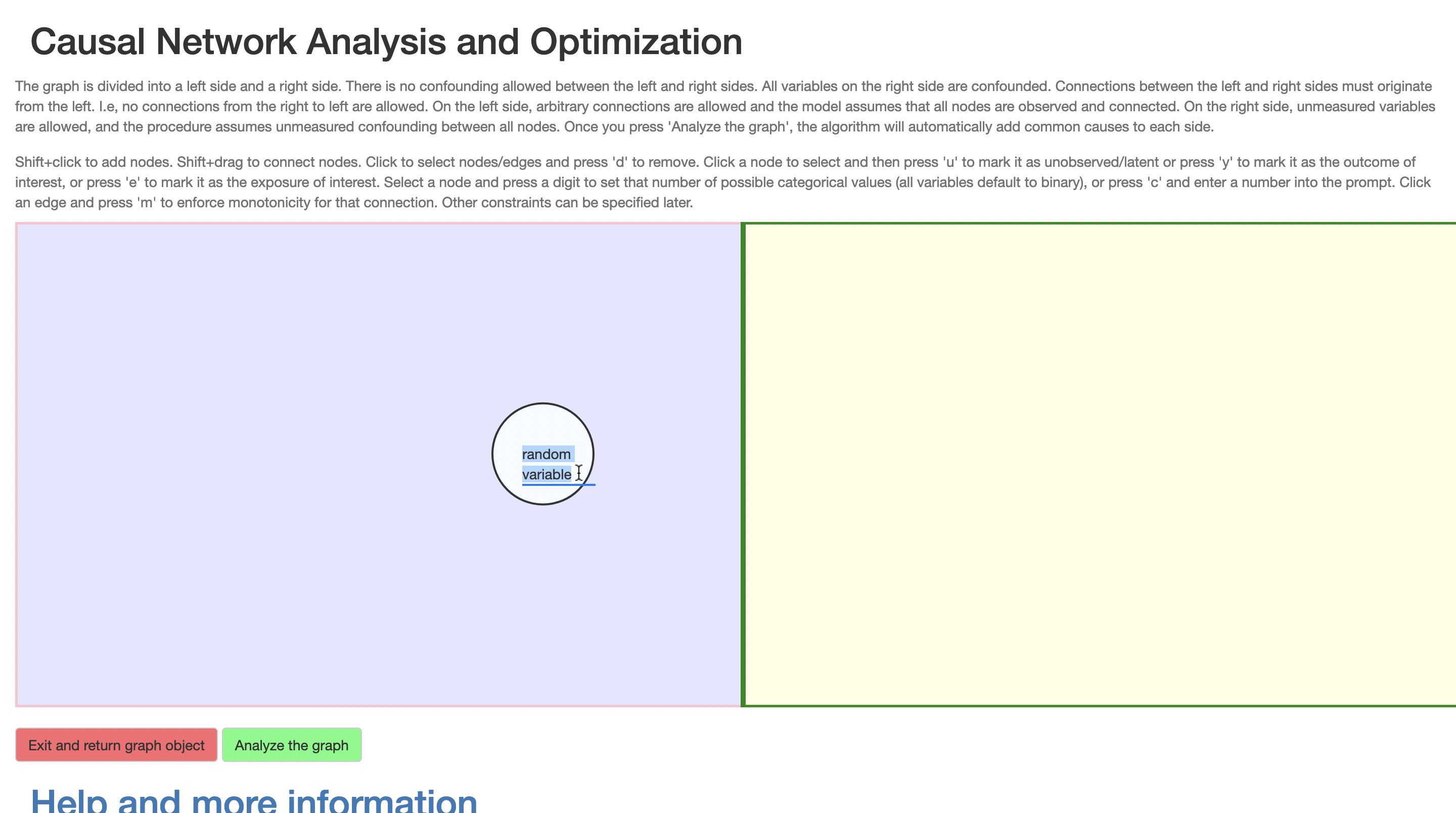} }\subfloat[Adding directed edges\label{fig:DAG-2}]{\includegraphics[width=0.5\linewidth]{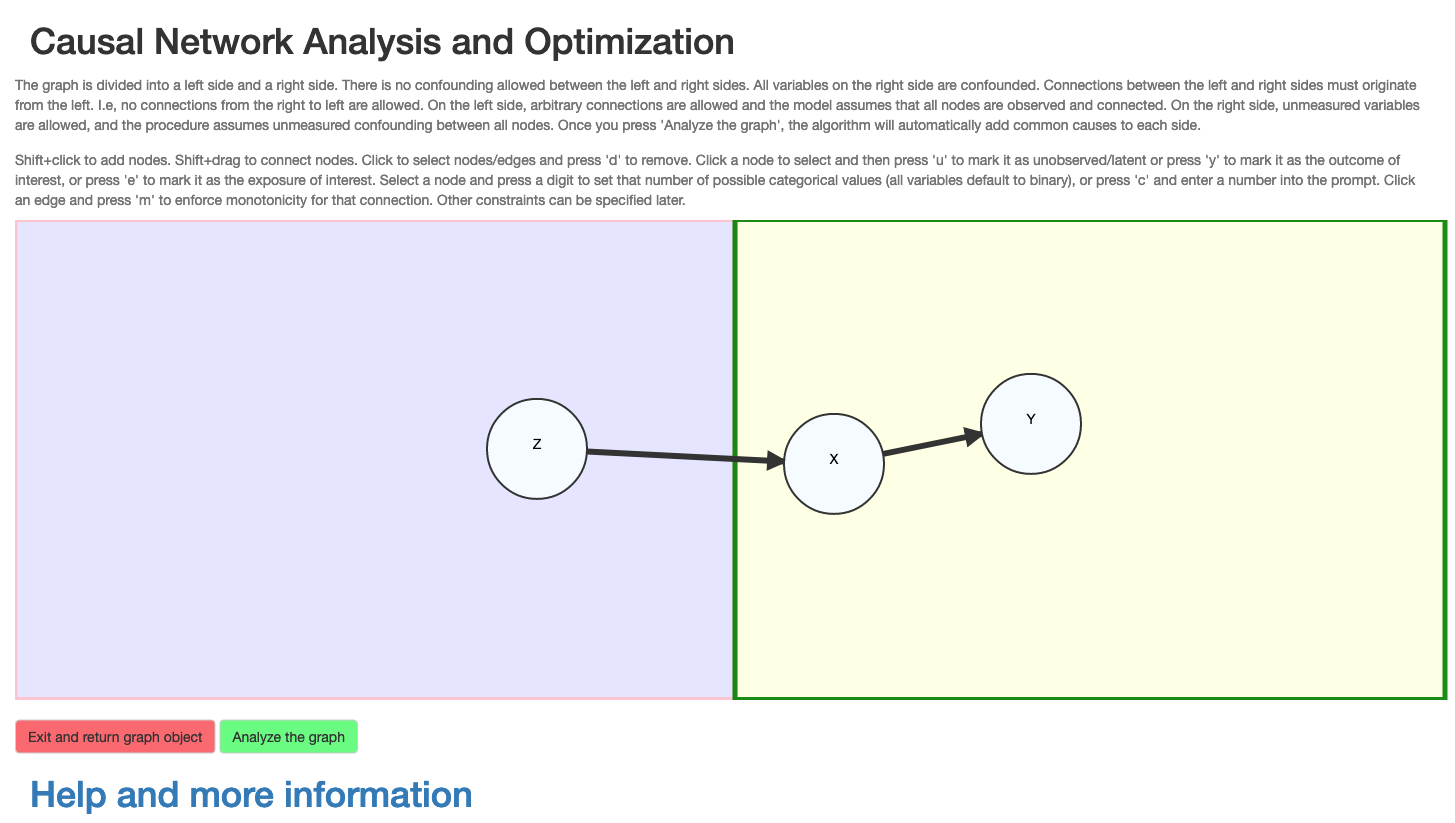} }\caption[Constructing the DAG]{Constructing the DAG}\label{fig:DAG}
\end{figure}
\end{Schunk}

The DAG is drawn using a point-and-click device (e.g., a mouse) to add
vertices representing variables (by Shift-click) and name them (using
any valid variable name in \texttt{R}), and to draw edges representing
direct causal influences (Shift+drag) between them. The vertices may
also be moved around, renamed and deleted (as can the edges) as also
described in an instruction text preceding the DAG interface. As shown
in Figure \ref{fig:DAG}, for the example we add a vertex \(Z\) on the
left side, and vertices \(X\) and \(Y\) on the right side. Then the
\(Z \to X\) and \(X \to Y\) edges are added by Shift+clicking on a
parent vertex and dragging to the child vertex. There is no need to add
the unmeasured confounder variable \(U_R\) as it is assumed and added
automatically.

Importantly, the nodes may be selected and assigned additional
information. In \(\mathcal{R}\) a variable may be assigned as unobserved
(click+`u'). All observed variables are assumed categorical and their
cardinality (i.e., number of levels) may be set (click+`c' brings up a
prompt for this this number; alternatively a short-cut click+`any digit'
is provided), with the default being binary. Although the causal query
(i.e., the causal effect of interest) is entered subsequently, the DAG
interface provides a convenient short-cut; a node \(X\) may be assigned
as an exposure (click+`e') and another \(Y\) as outcome (click+`y'),
whereupon the default query is the total causal risk difference
\(P(Y(X=1)=1)-P(Y(X=0)=1)\). Finally, an edge may be assigned as
representing an assumed monotonic influence (click+`m'). The nodes and
edges change appearance according to their assigned characteristics
(Figure \ref{fig:Cardinality}) and violations to the restrictions
characterizing the class of DAGs are detected and communicated to the
user.

\begin{Schunk}
\begin{figure}
\subfloat[Setting the number of categories\label{fig:Cardinality-1}]{\includegraphics[width=0.33\linewidth]{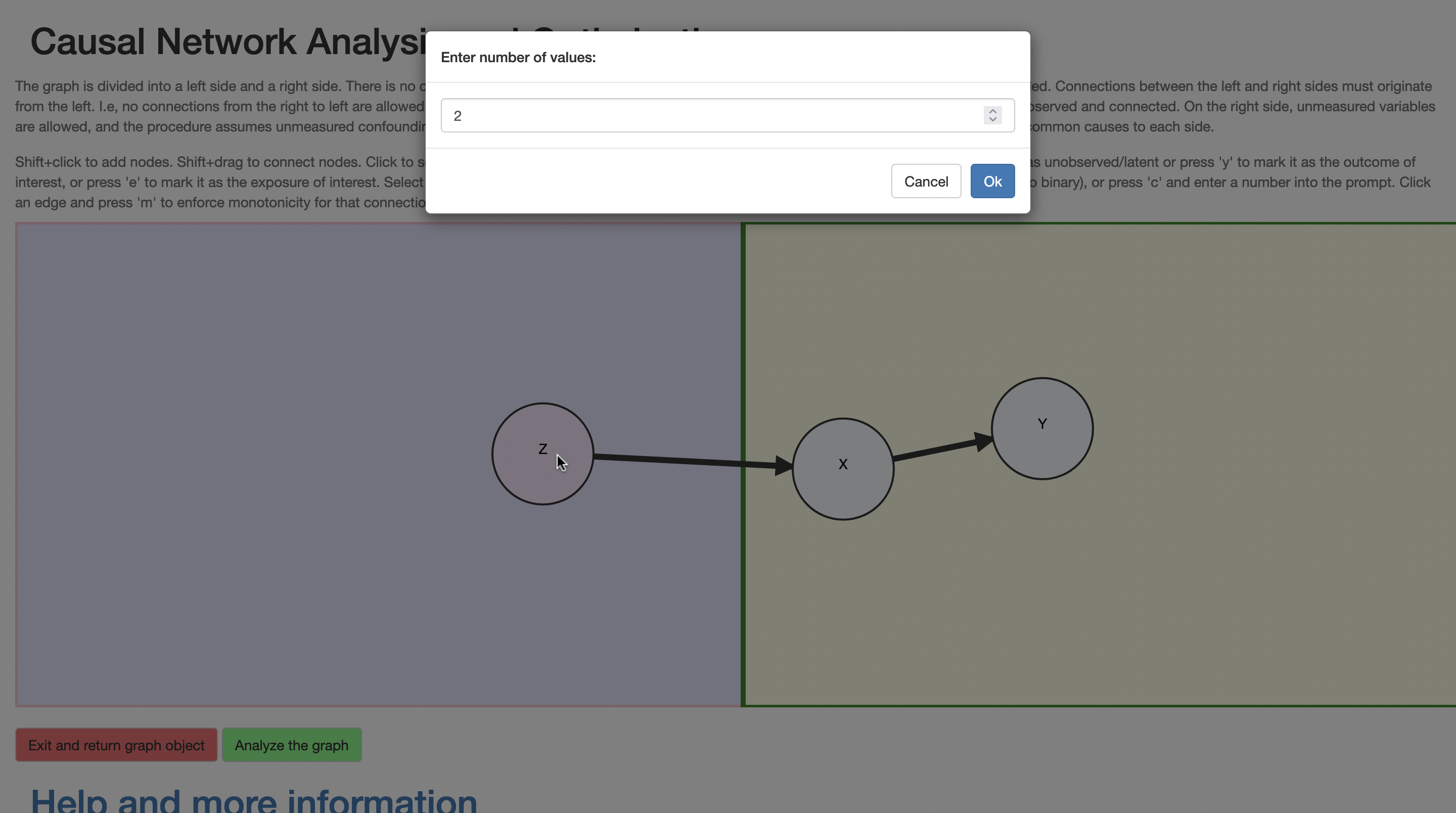} }\subfloat[Confirmation message\label{fig:Cardinality-2}]{\includegraphics[width=0.33\linewidth]{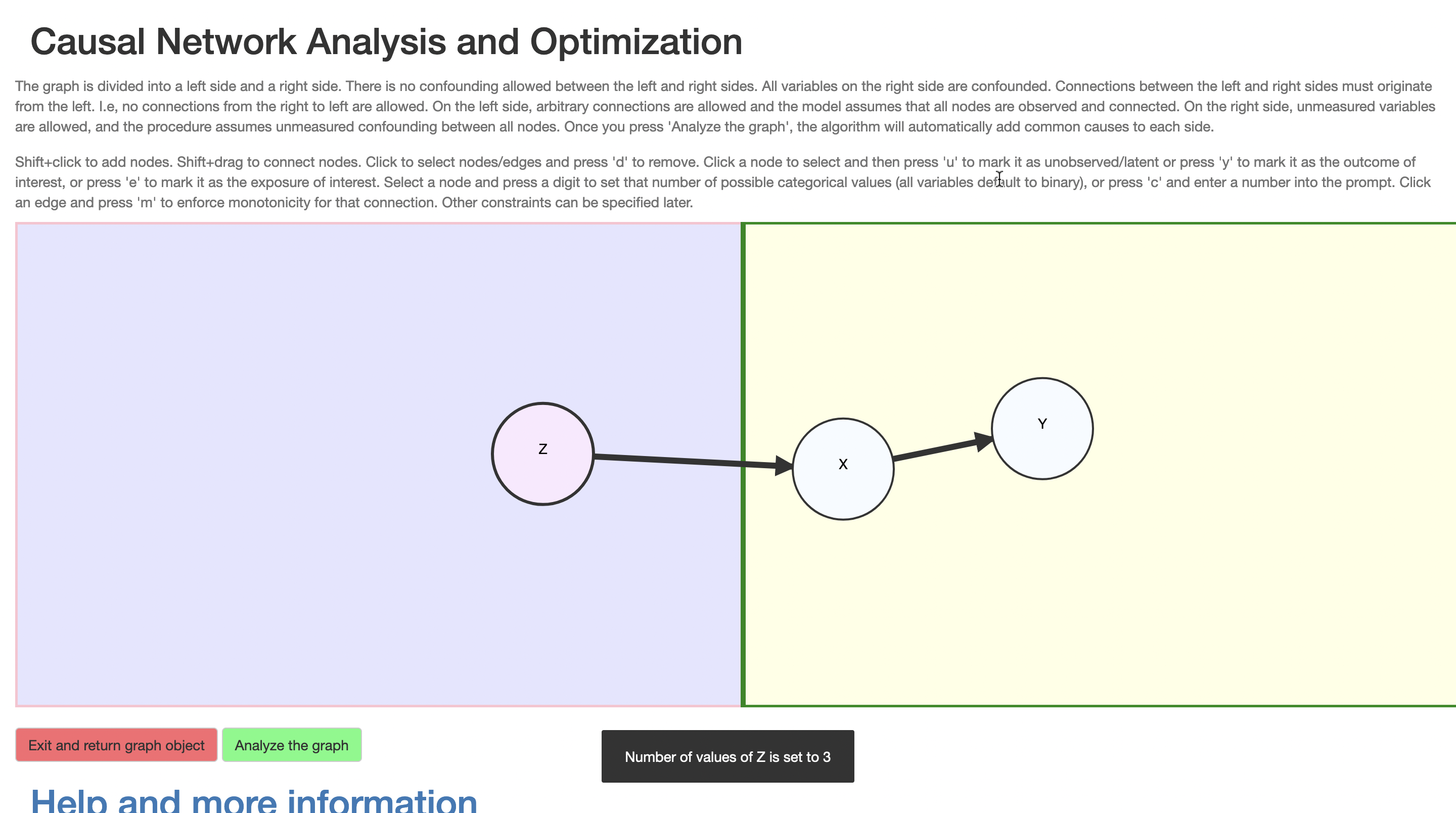} }\subfloat[Setting exposure and outcome\label{fig:Cardinality-3}]{\includegraphics[width=0.33\linewidth]{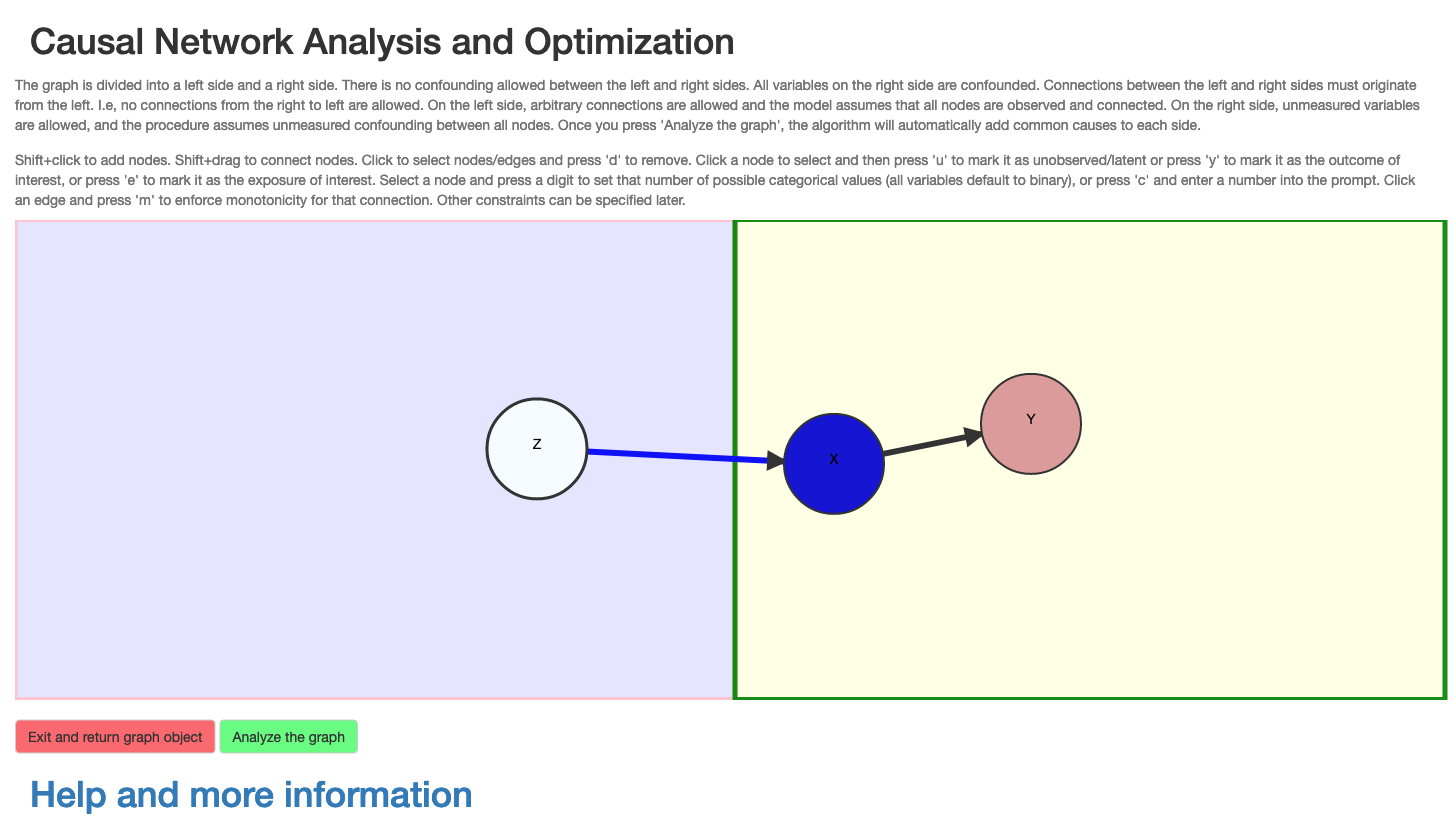} }\caption[Setting attributes]{Setting attributes}\label{fig:Cardinality}
\end{figure}
\end{Schunk}

Once the DAG has been drawn, the user may click the button ``Analyze the
graph'', upon which the DAG is interpreted and converted into an
annotated \CRANpkg{igraph}-object \citep{igraph} as described in the
implementation details below, and the results are displayed in graphical
form to the user (Figure \ref{fig:causalDAG}). The addition of \(U_R\),
the common unmeasured cause of \(X\) and \(Y\), is added and displayed
in this static plot.

\begin{Schunk}
\begin{figure}
\subfloat[Graphical summary of the DAG with added confounding\label{fig:causalDAG-1}]{\includegraphics[width=0.5\linewidth]{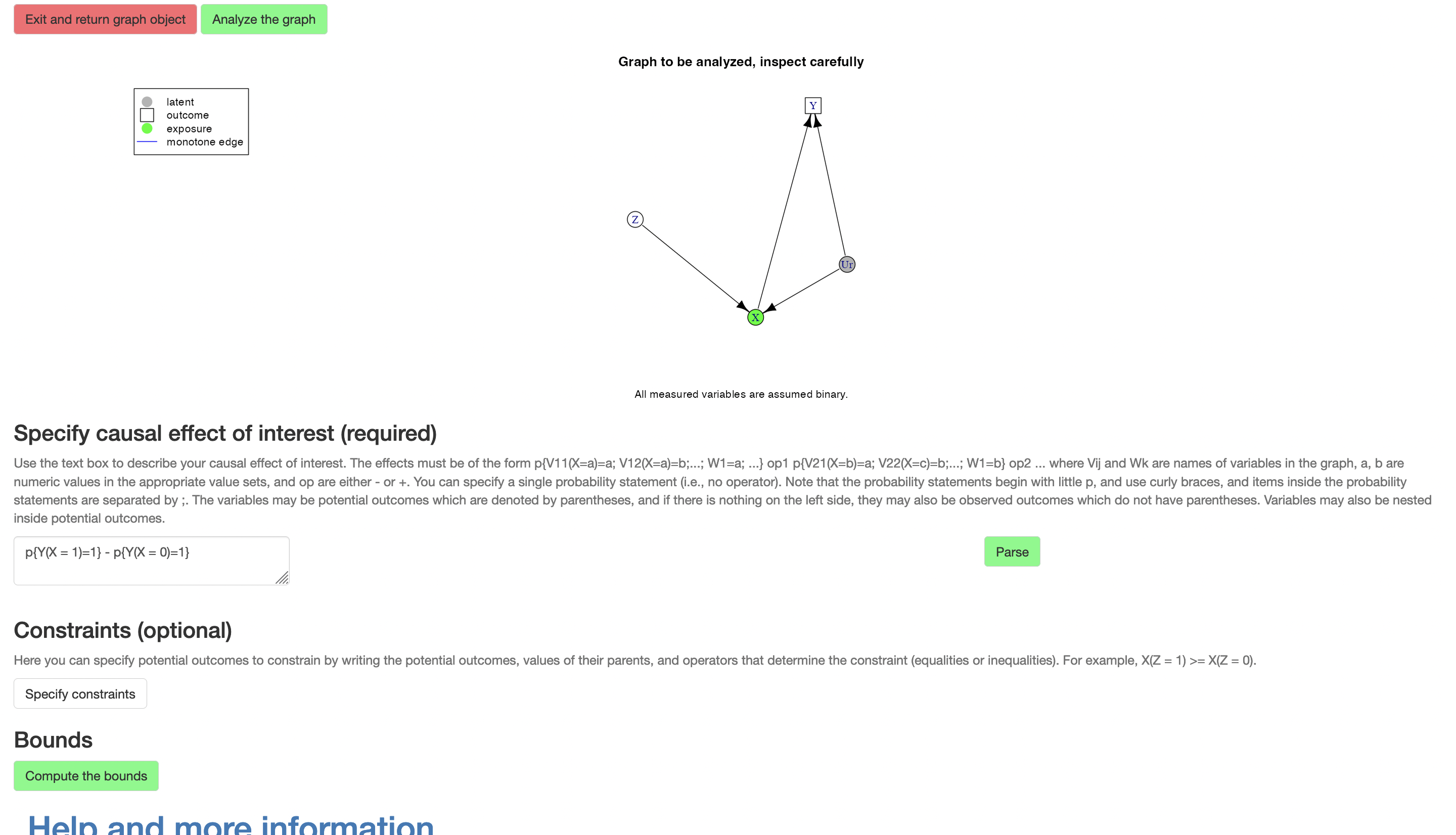} }\subfloat[Computing the bounds\label{fig:causalDAG-2}]{\includegraphics[width=0.5\linewidth]{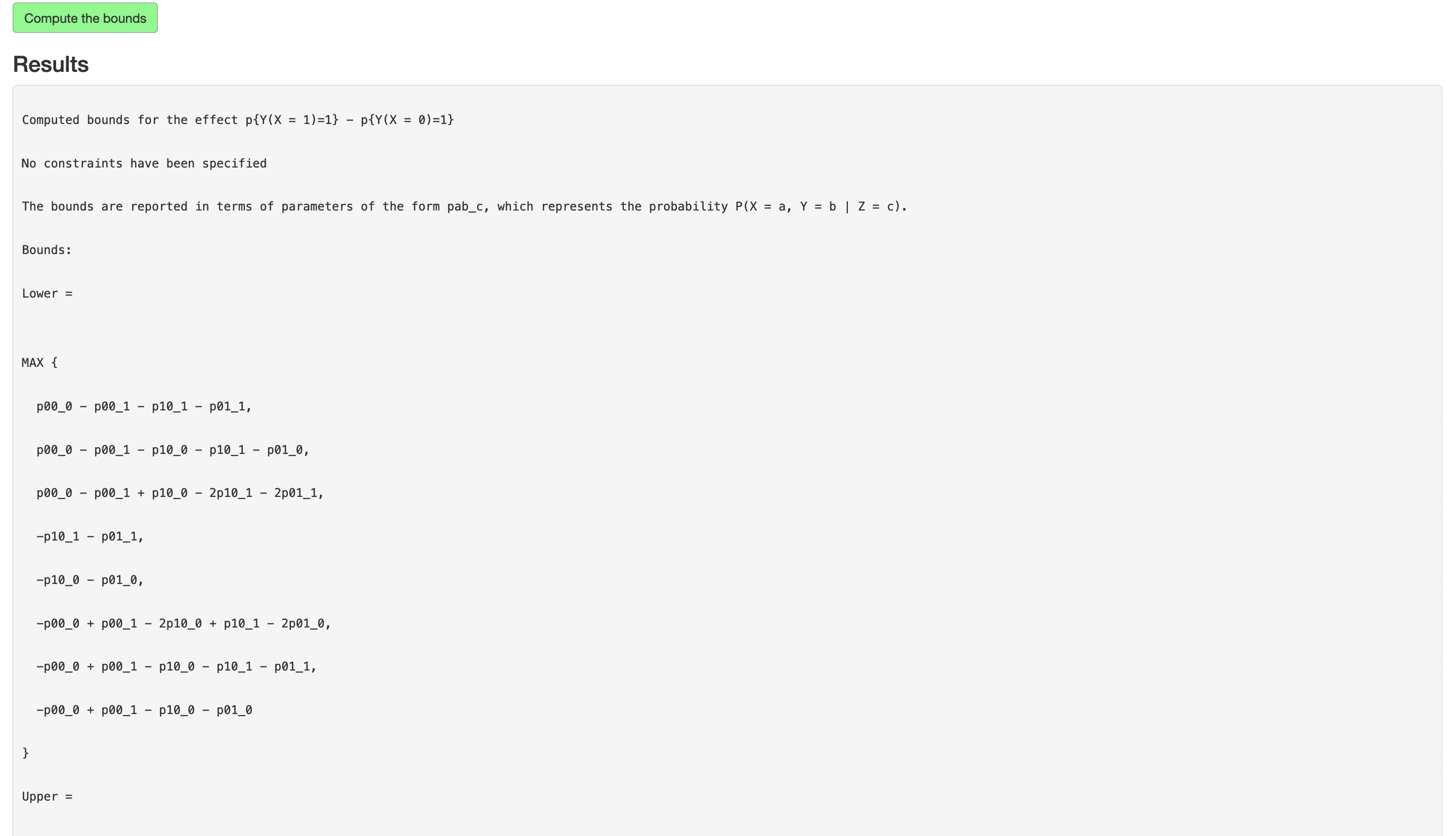} }\caption[The causal DAG and bounds]{The causal DAG and bounds}\label{fig:causalDAG}
\end{figure}
\end{Schunk}

\hypertarget{specifying-the-causal-query}{%
\subsection{Specifying the causal
query}\label{specifying-the-causal-query}}

Next, the user is asked to specify the causal query, i.e., causal effect
of interest. If no outcome variable has been assigned in the DAG then
the input field for the causal query is left blank and a query needs to
be specified. In our example, since we have assigned an exposure and
outcome using the DAG interface, the total causal risk difference
\(P(Y(X=1)=1)-P(Y(X=0)=1)\) is suggested.

\hypertarget{specifying-optional-additional-constraints}{%
\subsection{Specifying optional additional
constraints}\label{specifying-optional-additional-constraints}}

Finally, the user is given the option to provide any additional
constraints besides those imposed by the DAG. This may be considered an
optional advanced feature where, e.g., monotonicity of a certain direct
influence of \(Z\) on \(X\) may be assumed by entering
\(X(Z=1)\ge X(Z=0)\), with any such extra constraints separated by line
breaks. If this feature is used, the input is followed by clicking the
button ``Parse'', which identifies and fixes them.

\hypertarget{computing-the-symbolic-tight-bounds-on-the-query-under-the-given-constraints}{%
\subsection{Computing the symbolic tight bounds on the query under the
given
constraints}\label{computing-the-symbolic-tight-bounds-on-the-query-under-the-given-constraints}}

As the final step, the button ``Compute the bounds'' is clicked,
whereupon the constraints and objective are compiled into an
optimization problem which is then solved for tight causal bounds on the
query symbolically in terms of observational quantities (conditional
probabilities of the observed variables in the DAG) and the expressions
are displayed alongside information on how the parameters are to be
interpreted in terms of the given variable names (Figure
\ref{fig:causalDAG}). During computation, a progress indicator is shown,
and the user should be aware that complex and/or high-dimensional
problems may take significant time. The interface also provides a
feature to subsequently convert the bounds to \LaTeX-code using standard
probabilistic notation for publication purposes.

Once done, clicking ``Exit and return objects to R'' stops the
\CRANpkg{shiny} app and returns all information about the DAG, query and
computed bounds to the \texttt{R}-session. This information is bundled
in a list containing the graph, query, parameters and their
interpretation, the symbolic tight bounds as expressions as well as
implementations as \texttt{R}-functions and further log information
about the formulation and optimization procedures.

\hypertarget{programmatic-user-interface}{%
\section{Programmatic user
interface}\label{programmatic-user-interface}}

Interaction may also be done entirely programmatically as we illustrate
with the same binary instrumental variable example. First we create the
\texttt{igraph} object using the \texttt{graph\_from\_literal} function.
Once the basic graph is created, the necessary vertex and edge
attributes are added. The risk difference is defined as a character
object. The \texttt{analyze\_graph} function is the workhorse of
\CRANpkg{causaloptim}; it translates the causal graph, constraints, and
causal effect of interest into a linear programming problem. This linear
programming object, stored in \texttt{obj} in the code below, gets
passed to \texttt{optimize\_effect\_2} which performs vertex enumeration
to obtain the bounds as symbolic expressions in terms of observable
probabilities.

\begin{Schunk}
\begin{Sinput}
graph <- igraph::graph_from_literal(Z -+ X, X -+ Y, 
                                    Ul -+ Z, Ur -+ X, Ur -+ Y)
V(graph)$leftside <- c(1, 0, 0, 1, 0)
V(graph)$latent   <- c(0, 0, 0, 1, 1)
V(graph)$nvals    <- c(2, 2, 2, 2, 2)
E(graph)$rlconnect     <- c(0, 0, 0, 0, 0)
E(graph)$edge.monotone <- c(0, 0, 0, 0, 0)

riskdiff <- "p{Y(X = 1) = 1} - p{Y(X = 0) = 1}"
obj <- analyze_graph(graph, constraints = NULL, effectt = riskdiff)
bounds <- optimize_effect_2(obj)
bounds
\end{Sinput}
\begin{Soutput}
#> lower bound =  
#> MAX {
#>   p00_0 - p00_1 - p10_1 - p01_1,
#>   p00_0 - p00_1 - p10_0 - p10_1 - p01_0,
#>   p00_0 - p00_1 + p10_0 - 2p10_1 - 2p01_1,
#>   -p10_1 - p01_1,
#>   -p10_0 - p01_0,
#>   -p00_0 + p00_1 - 2p10_0 + p10_1 - 2p01_0,
#>   -p00_0 + p00_1 - p10_0 - p10_1 - p01_1,
#>   -p00_0 + p00_1 - p10_0 - p01_0
#> }
#> ----------------------------------------
#> upper bound =  
#> MIN {
#>   1 - p10_1 - p01_0,
#>   1 + p00_0 + p10_0 - 2p10_1 - p01_1,
#>   2 - p00_1 - p10_0 - p10_1 - 2p01_0,
#>   1 - p10_1 - p01_1,
#>   1 - p10_0 - p01_0,
#>   1 + p00_1 - 2p10_0 + p10_1 - p01_0,
#>   2 - p00_0 - p10_0 - p10_1 - 2p01_1,
#>   1 - p10_0 - p01_1
#> }
\end{Soutput}
\end{Schunk}

The resulting bounds object contains character strings representing the
bounds and logs containing detailed information from the vertex
enumeration algorithm. The bounds are printed to the console but more
features are available to facilitate their use. The
\texttt{interpret\_bounds} function takes the bounds and parameter names
as input and returns an \texttt{R} function implementing vectorized
forms of the symbolic expressions for the bounds.

\begin{Schunk}
\begin{Sinput}
bounds_function <- interpret_bounds(bounds$bounds, obj$parameters)
str(bounds_function)
\end{Sinput}
\begin{Soutput}
#> function (p00_0 = NULL, p00_1 = NULL, p10_0 = NULL, p10_1 = NULL, p01_0 = NULL, 
#>     p01_1 = NULL, p11_0 = NULL, p11_1 = NULL)
\end{Soutput}
\end{Schunk}

The results can also be used for numerical simulation using
\texttt{simulate\_bounds}. This function randomly generates
counterfactuals and probability distributions that satisfy the
constraints implied by the DAG and optional constraints. It then
computes and returns the bounds as well as the true causal effect.

If one wants to bound a different effect using the same causal graph,
the \texttt{update\_effect} function can be used to save some
computation time. It takes the object returned by
\texttt{analyze\_graph} and the new effect string then returns an object
of class \texttt{linearcausalproblem} that can be optimized:
\texttt{obj2\ \textless{}-\ update\_effect(obj,\ "p\{Y(X\ =\ 1)\ =\ 1\}")}.

Finally, \LaTeX-code may also be generated using the function
\texttt{latex\_bounds} as in
\texttt{latex\_bounds(bounds\$bounds,\ obj\$parameters)} yielding \tiny
\begin{align*}
 \mbox{Lower bound} &= \mbox{max} \left. \begin{cases}   P(X = 0, Y = 0 | Z = 0) - P(X = 0, Y = 0 | Z = 1) - P(X = 1, Y = 0 | Z = 1) - P(X = 0, Y = 1 | Z = 1),\\ 
   P(X = 0, Y = 0 | Z = 0) - P(X = 0, Y = 0 | Z = 1) - P(X = 1, Y = 0 | Z = 0) - P(X = 1, Y = 0 | Z = 1) - P(X = 0, Y = 1 | Z = 0),\\ 
   P(X = 0, Y = 0 | Z = 0) - P(X = 0, Y = 0 | Z = 1) + P(X = 1, Y = 0 | Z = 0) - 2P(X = 1, Y = 0 | Z = 1) - 2P(X = 0, Y = 1 | Z = 1),\\ 
   -P(X = 1, Y = 0 | Z = 1) - P(X = 0, Y = 1 | Z = 1),\\ 
   -P(X = 1, Y = 0 | Z = 0) - P(X = 0, Y = 1 | Z = 0),\\ 
   -P(X = 0, Y = 0 | Z = 0) + P(X = 0, Y = 0 | Z = 1) - 2P(X = 1, Y = 0 | Z = 0) + P(X = 1, Y = 0 | Z = 1) - 2P(X = 0, Y = 1 | Z = 0),\\ 
   -P(X = 0, Y = 0 | Z = 0) + P(X = 0, Y = 0 | Z = 1) - P(X = 1, Y = 0 | Z = 0) - P(X = 1, Y = 0 | Z = 1) - P(X = 0, Y = 1 | Z = 1),\\ 
   -P(X = 0, Y = 0 | Z = 0) + P(X = 0, Y = 0 | Z = 1) - P(X = 1, Y = 0 | Z = 0) - P(X = 0, Y = 1 | Z = 0) \end{cases} \right\} \\
 \mbox{Upper bound} &= \mbox{min} \left. \begin{cases}   1 - P(X = 1, Y = 0 | Z = 1) - P(X = 0, Y = 1 | Z = 0),\\ 
   1 + P(X = 0, Y = 0 | Z = 0) + P(X = 1, Y = 0 | Z = 0) - 2P(X = 1, Y = 0 | Z = 1) - P(X = 0, Y = 1 | Z = 1),\\ 
   2 - P(X = 0, Y = 0 | Z = 1) - P(X = 1, Y = 0 | Z = 0) - P(X = 1, Y = 0 | Z = 1) - 2P(X = 0, Y = 1 | Z = 0),\\ 
   1 - P(X = 1, Y = 0 | Z = 1) - P(X = 0, Y = 1 | Z = 1),\\ 
   1 - P(X = 1, Y = 0 | Z = 0) - P(X = 0, Y = 1 | Z = 0),\\ 
   1 + P(X = 0, Y = 0 | Z = 1) - 2P(X = 1, Y = 0 | Z = 0) + P(X = 1, Y = 0 | Z = 1) - P(X = 0, Y = 1 | Z = 0),\\ 
   2 - P(X = 0, Y = 0 | Z = 0) - P(X = 1, Y = 0 | Z = 0) - P(X = 1, Y = 0 | Z = 1) - 2P(X = 0, Y = 1 | Z = 1),\\ 
   1 - P(X = 1, Y = 0 | Z = 0) - P(X = 0, Y = 1 | Z = 1) \end{cases} \right\}.
 \end{align*} \normalsize

\hypertarget{implementation-and-program-overview}{%
\section{Implementation and Program
Overview}\label{implementation-and-program-overview}}

An overview of the main functions and their relations is depicted as a
flow chart in Figure \ref{fig:Overview}. All functions may be called
individually by the user at the \texttt{R}-console and all input, output
and interaction available through the \CRANpkg{shiny} app has
corresponding availability at the \texttt{R}-console as well.

\citep{generalcausalbounds} define the following class of problems for
which the query in general is not identifiable, but for which a
methodology to derive symbolic tight bounds on the query is provided.
The causal DAG consists of a finite set
\(\mathcal{W}=\{W_1,\dots,W_n\}=\mathcal{W}_\mathcal{L}\cup\mathcal{W}_\mathcal{R}\)
of categorical variables with
\(\mathcal{W}_\mathcal{L}\cap\mathcal{W}_\mathcal{R}=\varnothing\), no
edges going from \(\mathcal{W}_\mathcal{R}\) to
\(\mathcal{W}_\mathcal{L}\) and no external common parent between
\(\mathcal{W}_\mathcal{L}\) and \(\mathcal{W}_\mathcal{R}\), but
\emph{importantly} external common parents \(\mathcal{U}_\mathcal{L}\)
and \(\mathcal{U}_\mathcal{R}\) of variables within
\(\mathcal{W}_\mathcal{R}\) and \(\mathcal{W}_\mathcal{R}\) may not be
ruled out. Nothing is assumed about any characteristics of these
confounding variables \(\mathcal{U}_\mathcal{L}\) and
\(\mathcal{U}_\mathcal{R}\).

The causal query may be any linear combination of joint probabilities of
factual and counterfactual outcomes expressed in terms of the variables
in \(\mathcal{W}\) and may always be expressed as a sum of probabilities
of response function variables of the DAG. It is subject to the
restriction that each outcome variable is in \(\mathcal{W}_\mathcal{R}\)
and if \(\mathcal{W}_\mathcal{L}\ne\varnothing\) it is also subject to a
few regularity conditions as detailed in \citep{generalcausalbounds}.
Tight bounds on the query may then be derived symbolically in terms of
conditional probabilities of the observable variables \(\mathcal{W}\).

Algorithms 1 and 2 in \citep{generalcausalbounds} construct the
constraint space and causal query in terms of the joint probabilities of
the response function variables and in \CRANpkg{causaloptim} are
implemented in the functions \texttt{create\_R\_matrix} and
\texttt{create\_effect\_vector} respectively. Both are called as
sub-procedures of the function \texttt{analyze\_graph} to translate the
causal problem to that of optimizing an objective function over a
constraint space. The implementation of Algorithm 1 involves
constructing the response functions themselves as actual
\texttt{R}-functions. Evaluating these correspond to evaluating the
structural equations of the causal DAG.

The conditions on the DAG suffice to ensure that the causal query will
depend only on the response functions corresponding to the variables in
\(\mathcal{W}_\mathcal{R}\) and that the exhaustive set of constraints
of their probabilities are linear in a subset of conditional
probabilities of observable variables (Proposition 2 in
\citep{generalcausalbounds}), and the conditions on the query in turn
ensure that it may be expressed as a linear combination of joint
probabilities of the response functions of the variables in
\(\mathcal{W}_\mathcal{R}\) (Proposition 3 in
\citep{generalcausalbounds}).

Once this formulation of the causal problem as a linear program has been
set up, a vertex enumeration method is employed to compute the extrema
symbolically in terms of conditional probabilities of the observable
variables.

The main and interesting functions will be described in some detail
below. We begin however with an overview of how they are tied together
by the \CRANpkg{shiny} app.

\begin{Schunk}
\begin{figure}

{\centering \includegraphics[width=1\linewidth]{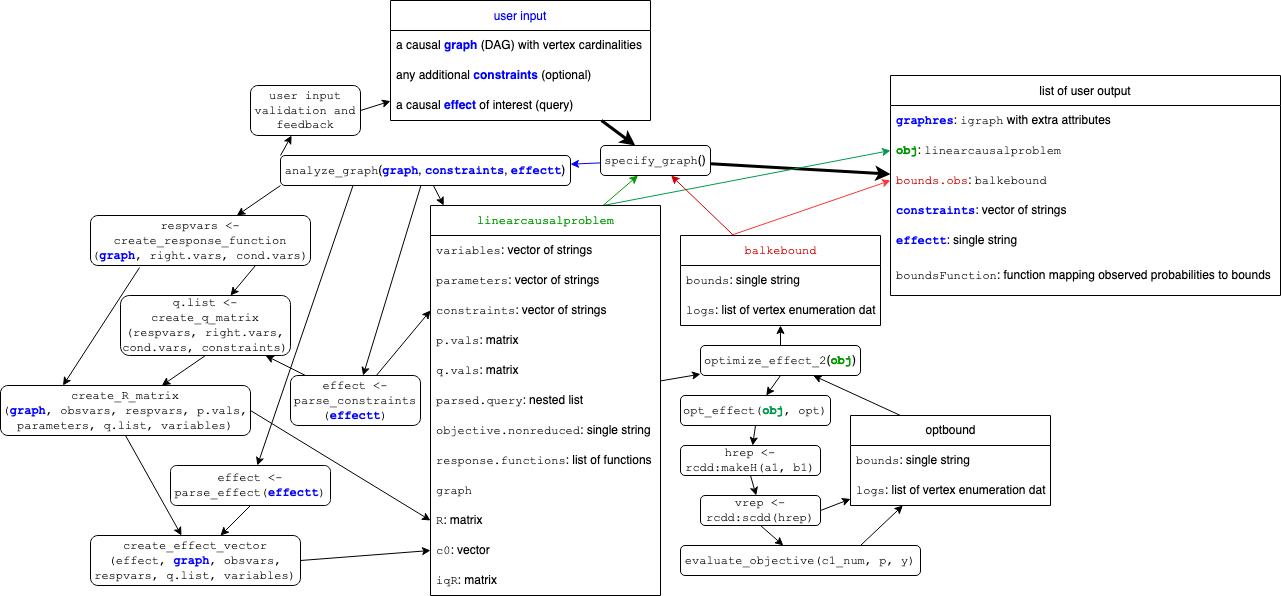} 

}

\caption[Function Overview Flow Chart]{Function Overview Flow Chart}\label{fig:Overview}
\end{figure}
\end{Schunk}

\hypertarget{specify_graph}{%
\subsubsection{\texorpdfstring{\texttt{specify\_graph}}{specify\_graph}}\label{specify_graph}}

The graphical interface is launched by \texttt{specify\_graph()}, or
preferably \texttt{results\ \textless{}-\ specify\_graph()}. Once the
\CRANpkg{shiny} app is stopped, the input, output and other useful
information is returned by the function, so we recommend assigning it to
a variable so they are saved in the \texttt{R}-session and may easily be
further analyzed and processed. All further function calls will take
place automatically as the user interacts with the web interface. Thus,
from a basic user perspective, \texttt{specify\_graph} is the main
function. The core functionality however is implemented in the functions
\texttt{analyze\_graph}, \texttt{optimize\_effect\_2} and their
subroutines.

The \texttt{JavaScript} that handles the communication between the
\CRANpkg{shiny} server and the input as the user draws a DAG through the
web interface uses on the project \texttt{directed-graph-creator}, an
interactive tool for creating directed graphs, created using
\texttt{d3.js} and hosted at
\url{https://github.com/cjrd/directed-graph-creator}, which has been
modified for the purpose of causal diagrams. The modification binds the
user inputs as they interact with the graph to \CRANpkg{shiny} so that
the directed graph and its attributes set by the user are reactively
converted into an \texttt{igraph}-object for further processing. Since
directed graphs are common in many computational and statistical
problems, this \CRANpkg{shiny} interface may also be valuable to many
other \texttt{R}-package authors and maintainers who may wish to provide
their users with an accessible and intuitive way to interact with their
software.

The server listens to a reactive function that, as the user draws the
DAG, collects information about the current edges, collects and
annotates vertices, adds left- and right-side confounding, and returns
an annotated \texttt{igraph}-object, comprising information about the
connectivity along with some additional attributes; for each variable,
its name, cardinality, latency-indicator and side-indicator, and for
each edge, a monotonicity-indicator and (to detect and communicate
violations on direction) a right-to-left-indicator. The server meanwhile
also monitors the DAG for any violation of the restriction that each
edge between \(\mathcal{L}\) and \(\mathcal{R}\) must go \emph{from}
\(\mathcal{L}\) \emph{to} \(\mathcal{R}\), and if detected directly
communicates this to the user through a text message in the
\CRANpkg{shiny} app.

\hypertarget{analyze_graph}{%
\subsubsection{\texorpdfstring{\texttt{analyze\_graph}}{analyze\_graph}}\label{analyze_graph}}

The function \texttt{analyze\_graph} takes a DAG (in the form of an
\texttt{igraph} object), optional constraints, and a string representing
the causal effect of interest and proceeds to construct and return a
linear optimization problem (a \texttt{linearcausalproblem}-object) from
these inputs.

First, some basic data structures are created to keep track of the
observed variables, their possible values, the latent variables, and
whether they are in \(\mathcal{L}\) or \(\mathcal{R}\). Once these basic
data-structures have been created, the first task of the algorithm is to
create the response function variables (for each variable, observed or
not, except \(U_\mathcal{L}\) and \(U_\mathcal{R}\)). Probabilities of
these will be the entities \(\mathbf{q}\) in which the objective
function (representing the target causal effect) is expressed and will
constitute the points in the space it is optimized over, where this
space itself is constrained by the the relationships between them and
observed conditional probabilities \(\mathbf{p}\).

\hypertarget{create_response_function}{%
\subsubsection{\texorpdfstring{\texttt{create\_response\_function}}{create\_response\_function}}\label{create_response_function}}

The function \texttt{create\_response\_function} returns a list
\texttt{respvars} that has a named entry for each observed variable,
containing its response function variable and response function. If
\(X\) is an observed variable with \(n\) response functions, then they
are enumerated by \(\{0,\dots,n-1\}\). Its entry \texttt{respvars\$X}
contains the response function variable \(R_X\) of \(X\), and is a list
with two entries. The first, \texttt{respvars\$X\$index}, is a vector
containing all the possible values of \(R_X\), i.e., the integers
\((0,\dots,n-1)\). The second, \texttt{respvars\$X\$values} is itself a
list with \(n\) entries; each containing the particular response
function of \(X\) corresponding to its index. Each such response
function is an actual \texttt{R}-function and may be evaluated by
passing it any possible values of the parents of \(X\) as arguments.

Next, the response function variables are used in the creation of a
matrix of unobserved probabilities. Specifically the joint probabilities
\(P(\mathbf{R}_\mathcal{R}=\mathbf{r}_\mathcal{R})\) for each possible
value-combination \(\mathbf{r}_\mathcal{R}\) of the response function
variables \(\mathbf{R}_\mathcal{R}\) of the right-side-variables
\(\mathbf{W}_\mathcal{R}\). In \citep{generalcausalbounds}, the possible
value-combinations \(\mathbf{r}_\mathcal{R}\) are enumerated by
\(\gamma\in\{1,\dots,\aleph_\mathcal{R}\}\) with corresponding
probabilities \(q_\gamma:=P(\mathbf{R}_\mathcal{R}=\mathbf{r}_\gamma)\)
being components of the vector
\(\mathbf{q}\in[0,1]^{\aleph_\mathcal{R}}\).

\hypertarget{create_r_matrix}{%
\subsubsection{\texorpdfstring{\texttt{create\_R\_matrix}}{create\_R\_matrix}}\label{create_r_matrix}}

The constraints that the DAG and observed conditional probabilities
\(\mathbf{p}\) (in \texttt{p.vals}) impose on the unobserved
probabilities \(\mathbf{q}\) (represented by \texttt{variables}) are
linear. Specifically, there exists a matrix whose entries are the
coefficients relating \texttt{p.vals} to \texttt{variables}. This matrix
is called \(P\) in \citep{generalcausalbounds}, where its existence is
guaranteed by Proposition 2 and its construction is detailed in
Algorithm 1, which is implemented in the function
\texttt{create\_R\_matrix}. This function returns back a list with two
entries; a vector of strings representing the linear constraints on the
unobserved \(\mathbf{q}\in[0,1]^{\aleph_\mathcal{R}}\) imposed by and in
terms of the observed \(\mathbf{p}\in[0,1]^B\) and the numeric matrix
\(R\in\{0,1\}^{(B+1)\times\aleph_\mathcal{R}}\) of coefficients
corresponding to these constraints as well as the probabilistic ones and
given by \(R=\begin{pmatrix}\mathbf{1}\\P\end{pmatrix}\) where
\(P\in\{0,1\}^{B\times\aleph_\mathcal{R}}:\mathbf{p}=P\mathbf{q}\), so
\(R\mathbf{q}=\begin{pmatrix}1\\\mathbf{p}\end{pmatrix}\).

This determines the constraint space as a compact convex polytope in
\(\mathbf{q}\)-space, i.e., in \(\mathbb{R}^{\aleph_\mathcal{R}}\). To
create the matrix, we define a recursive function \texttt{gee\_r} that
takes two arguments; a positive integer \texttt{i} being the index
\(i\in\{1,\dots,n\}\) of a variable \(W_i\in\mathcal{W}\) (i.e.~the
\(i^{th}\) component of \(\mathbf{W}\) or, equivalently, the
\texttt{i}th entry of \texttt{obsvars}) and a vector \texttt{r} being a
value \(\mathbf{r}\in\nu(\mathbf{R})\) in the set \(\nu(\mathbf{R})\) of
all possible value-vectors of the joint response function variable
\(\mathbf{R}\). This recursive function is called for each variable in
\texttt{obsvars} and for each possible value of the response function
variable vector. The base case is reached if the variable has no
parents, in which case the list corresponding to the response function
variable \(R_{W_i}\) of \(W_i\) is extracted from \texttt{respvars}.
From this list, the entry whose index matches the \texttt{i}th index of
\texttt{r} (i.e.~the one corresponding to the response function variable
value \(r_i=\)\texttt{r{[}i{]}}) is extracted and finally its value,
i.e., the corresponding response function itself, is extracted and is
evaluated on an empty list of arguments, since it is a constant function
and determined only by the value \(r_i\).

The recursive case is encountered when \texttt{parents} is non-empty. If
so, then for each parent in \texttt{parents}, its index in
\texttt{obsvars} is determined and \texttt{gee\_r} is recursively called
with the same vector \texttt{r} as first argument but now with this
particular index (i.e.~that of the current parent) as second argument.
The numeric values returned by these recursive calls are then
sequentially stored in a vector \texttt{lookin}, whose entries are named
by those in \texttt{parents}. Just as in the base case, the response
function corresponding to the particular value \(r_i\) of the response
function variable \(R_{W_i}\) (i.e.~the response function of the
variable \texttt{obsvars{[}i{]}} that has the index \texttt{r{[}i{]}})
is extracted from \texttt{respvars} and is now evaluated with arguments
given by the list \texttt{lookin}. Note that \texttt{gee\_r(r,\ i)}
corresponds to the value \(w_i=g^*_{W_i}(\mathbf{r})\) in
\citep{generalcausalbounds}.

Then the values that match the observed probabilities are recorded, the
corresponding entries in the current row of the matrix \texttt{R} are
set to 1 and a string representing the corresponding equation is
constructed and added to the vector of constraints.

\hypertarget{parse_effect}{%
\subsubsection{\texorpdfstring{\texttt{parse\_effect}}{parse\_effect}}\label{parse_effect}}

Now that the constraint space has been determined, the objective
function representing the causal query needs to be specified as a linear
function of the components of \(\mathbf{q}\), i.e., \texttt{variables}.
First the causal query that has been provided by the user as a
text-string in \texttt{effectt} is passed to the function
\texttt{parse\_effect}, which identifies its components including nested
counterfactuals and creates a data structure representing the causal
query. This structure includes nested lists which represent all
interventional paths to each outcome variable in the query.

Once the nested list \texttt{effect} is returned back to
\texttt{analyze\_graph}, it checks that the requirements (see
Proposition 3 in \citep{generalcausalbounds}) on the query are fulfilled
before creating the linear objective function. Despite these regularity
conditions, a large set of possible queries may be entered using
standard counterfactual notation, using syntax described in the
accompanying instruction text along with examples such as
\(P(Y(M(X = 0), X = 1) = 1) - P(Y(M(X = 0), X = 0) = 1)\); the natural
direct effect \citep{pearl2001direct} of a binary exposure \(X\) at
level \(M=0\) on a binary outcome \(Y\) \emph{not} going through the
mediator \(M\), in the presence of unmeasured confounding between \(M\)
and \(Y\) \citep{sjolander2009bounds}.

\hypertarget{create_effect_vector}{%
\subsubsection{\texorpdfstring{\texttt{create\_effect\_vector}}{create\_effect\_vector}}\label{create_effect_vector}}

Now that the required characteristics of the query have been
established, the corresponding objective function will be constructed by
the function \texttt{create\_effect\_vector} which returns a list
\texttt{var.eff} of string-vectors; one for each term in the query. Each
such vector contains the names (strings in \texttt{variables}) of the
response function variables of the right-side (i.e.~the components of
\(\mathbf{q}\)) whose sum corresponds the that particular term. The
function \texttt{create\_effect\_vector} implements Algorithm 2 of
\citep{generalcausalbounds} with the additional feature that if the user
has entered a query that is incomplete in the sense that there are
omitted mediating variables on paths from base/intervention variables to
the outcome variable, then this is interpreted as the user intending the
effects of the base/intervention variables to be propagated through the
mediators, so that they are set to their ``natural'' values under this
intervention. These mediators are detected and their values are set
accordingly.

We define a recursive function \texttt{gee\_rA} that takes three
arguments; a positive integer \texttt{i} (the index \(i\) of a variable
\(W_i\in\mathcal{W}=\)\texttt{obsvars}), a vector \texttt{r} (a value
\(\mathbf{r}\in\nu(\mathbf{R})\) in the set \(\nu(\mathbf{R})\) of all
possible value-vectors of the joint response function variable
\(\mathbf{R}\)) and a string \texttt{path} that represents an
interventional path and is of the form ``X -\textgreater{} \ldots{}
-\textgreater{} Y'' if not \texttt{NULL}. The base case is reached
either if \texttt{path} is non-\texttt{NULL} and corresponds to a path
to the intervention set or if \texttt{parents} is empty. In the former
case, the corresponding numeric intervention-value is returned, and in
the latter case, the value of the corresponding response function called
on the empty list of arguments is returned just as in the base case of
\texttt{gee\_r}. The recursive case is encountered when \texttt{path} is
\texttt{NULL} and \texttt{parents} is non-empty. This recursion proceeds
just as in \texttt{gee\_r}, but now rather with a recursive call to
\texttt{gee\_rA}, whose third argument is now
\texttt{path\ =\ paste(gu,\ "-\textgreater{}",\ path)} where the string
in \texttt{gu} is the name of the parent variable in \texttt{parents}
whose index \texttt{i} in \texttt{obsvars} is the second argument of
this recursive call. This construction traces the full path taken from
the outcome of interest to the variable being intervened upon. Note that
\texttt{gee\_rA(r,\ i,\ path)} corresponds to the value
\(w_i=h^{A_i}_{W_i}(\mathbf{r},W_i)\) in \citep{generalcausalbounds}. A
matrix is now created just as in the observational case, but this time
using \texttt{gee\_rA} instead of \texttt{gee\_r} .

\hypertarget{optimize_effect_2}{%
\subsubsection{\texorpdfstring{\texttt{optimize\_effect\_2}}{optimize\_effect\_2}}\label{optimize_effect_2}}

Once the constraints on \(\mathbf{q}\) as well as the effect of interest
in terms of \(\mathbf{q}\) have been established, it remains only to
optimize this expression over the constraint space. Here, \(\mathbf{c}\)
denotes the constant gradient vector of the linear objective function
and \(P\) denotes the coefficient matrix of the linear restrictions on
\(\mathbf{q}\) in terms of \(\mathbf{q}\) imposed by the causal DAG. By
adding the probabilistic constraints on \(\mathbf{q}\) we have arrived
at e.g.~the following linear program giving a tight lower bound on the
average causal effect
\(\theta_\mathbf{q} = P\{Y(X = 1) = 1\} - P\{Y(X = 0) = 1\}\) in the
simple instrumental variable problem of the introductory section:

\begin{align*}
\min_\mathbf{q} \theta_\mathbf{q}
&=\min\{\mathbf{c}^\top\mathbf{q}\mid \mathbf{q}\in\mathbb{R}^{16},\mathbf{q}\geq\mathbf{0}_{16\times1},\mathbf{1}_{1\times 16} \mathbf{q}=1,P\mathbf{q}=\mathbf{q}\}\\
&=\max\{\begin{pmatrix}1&\mathbf{q}^\top\end{pmatrix}\mathbf{y}\mid\mathbf{y}\in\mathbb{R}^{9},\mathbf{y}\geq \mathbf{0}_{9\times 1},\begin{pmatrix}\mathbf{1}_{16\times 1}&P^\top\end{pmatrix}\leq\mathbf{c}\}\\
&=\max\{\begin{pmatrix}1&\mathbf{q}^\top\end{pmatrix}\bar{\mathbf{y}}\mid\bar{\mathbf{y}}\text{ is a vertex of }\{\mathbf{y}\in\mathbb{R}^{9}\mid \mathbf{y}\geq\mathbf{0}_{9\times 1},R^\top\leq\mathbf{c}\}\}
\end{align*}

Since we allow the user to provide additional linear inequality
constraints (e.g.~it may be quite reasonable to assume the proportion of
``defiers'' in the study population of our example to be quite low), the
actual primal and dual linear programs may look slightly more
complicated, but this small example still captures the essentials.\\
In general, given the matrix of linear constraints on the observable
probabilities implied by the DAG and an optional user-provided matrix
inequality, we construct the coefficient matrix and right hand side
vector of the dual polytope.

The optimization via vertex enumeration step in \CRANpkg{causaloptim} is
implemented in the function \texttt{optimize\_effect\_2} which uses the
double description method for vertex enumeration, as implemented in the
\CRANpkg{rcdd} package \citep{rcdd}. This step of vertex enumeration has
previously been the major computational bottleneck. The approach is now
based on \texttt{cddlib}
(\url{https://people.inf.ethz.ch/fukudak/cdd_home/}), which has an
implementation of the Double Description Method (dd). Any convex
polytope can be dually described as either an intersection of
half-planes (which is the form we get our dual constraint space in) or
as a minimal set of vertices of which it is the convex hull (which is
the form we want it in) and the dd algorithm efficiently converts
between these two descriptions. \texttt{cddlib} also uses exact rational
arithmetic, so there is no need to worry about any numerical instability
issues. The vertices of the dual polytope are obtained and stored as
rows of a matrix with \texttt{hrep\ \textless{}-\ rcdd::makeH(a1,\ b1);}
\texttt{vrep\ \textless{}-\ rcdd::scdd(hrep);}
\texttt{vertices\ \textless{}-\ vrep\$output{[}vrep\$output{[},\ 1{]}\ ==\ 0\ \&\ vrep\$output{[},\ 2{]}\ ==\ 1,\ -c(1,\ 2),\ drop=FALSE{]}}.

The rest is simply a matter of plugging them into the dual objective
function, evaluating the expression and presenting the results. The
first part of this is done by
\texttt{apply(vertices,\ 1,\ function(y)\ evaluate\_objective(c1\_num,\ p,\ y))}
(here
\texttt{(c1\_num,p)}\(=(\begin{pmatrix}b_{\ell}^\top&1\end{pmatrix},p)\)
separates the dual objective gradient into its numeric and symbolic
parts).

\CRANpkg{causaloptim} also contains a precursor to to
\texttt{optimize\_effect\_2}, called \texttt{optimize\_effect}. This
legacy function uses the original optimization procedure written in
\texttt{C++} by Alexander Balke and involves linear program formulation
followed by the vertex enumeration algorithm of
\citep{mattheiss1973algorithm}. This has worked well for very simple
settings but has struggled severely with even remotely complex ones and
thus been insufficient for the ambitions of \CRANpkg{causaloptim}. The
efficiency gains of \texttt{optimize\_effect\_2} over the legacy code
have reduced the computation time for several setting from hours to
milliseconds.

\hypertarget{numeric-examples}{%
\section{Numeric Examples}\label{numeric-examples}}

\hypertarget{a-mediation-analysis}{%
\subsection{A Mediation Analysis}\label{a-mediation-analysis}}

In \citep{sjolandernaturaldirecteffects}, the author derives bounds on
natural direct effects in the presence of confounded intermediate
variables and applies them to data from the Lipid Research Clinics
Coronary Primary Prevention Trial \citep{freedmandata}, where subjects
were randomized to cholestyramine treatment and presence of coronary
heart disease events as well as levels of cholesterol were recorded
after a 1-year follow-up period. We let \(X\) be a binary treatment
indicator, with \(X=0\) indicating actual cholestyramine treatment and
\(X=1\) indicating placebo. We further let \(Y\) be an indicator of the
occurrence of coronary heart disease events within follow-up, with
\(Y=0\) indicating event-free follow-up and \(Y=1\) indicating an event.
We finally let \(M\) be a dichotomized (cut-off at \(280\ mg/dl\))
cholesterol level indicator, with \(M=0\) indicating levels
\(<280\ mg/dl\) and \(M=1\) indicating levels \(\ge280\ mg/dl\). The
causal assumptions are summarized in the DAG shown in Figure
\ref{fig:mediation-fig}, where \(U_l\) and \(U_r\) are unmeasured and
the latter confounds the effect of \(M\) on \(Y\).

\begin{Schunk}
\begin{figure}

{\centering \includegraphics[width=1\linewidth]{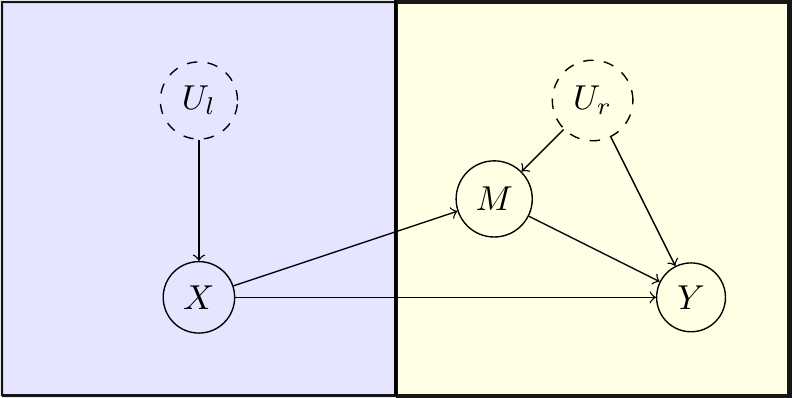} 

}

\caption[Causal DAG for mediation example]{Causal DAG for mediation example}\label{fig:mediation-fig}
\end{figure}
\end{Schunk}

\begin{Schunk}
\begin{Sinput}
b <- igraph::graph_from_literal(X -+ Y, X -+ M, M -+ Y, 
                                Ul -+ X, Ur -+ Y, Ur -+ M)
V(b)$leftside <- c(1, 0, 0, 1, 0)
V(b)$latent   <- c(0, 0, 0, 1, 1)
V(b)$nvals    <- c(2, 2, 2, 2, 2)
E(b)$rlconnect     <- c(0, 0, 0, 0, 0, 0)
E(b)$edge.monotone <- c(0, 0, 0, 0, 0, 0)
\end{Sinput}
\end{Schunk}

Using the data from Table IV of \citep{sjolandernaturaldirecteffects},
we compute the observed conditional probabilities.

\begin{Schunk}
\begin{Sinput}
# parameters of the form pab_c, which represents 
# the probability P(Y = a, M = b | X = c)
p00_0 <- 1426/1888 # P(Y=0,M=0|X=0)
p10_0 <- 97/1888 # P(Y=1,M=0|X=0)
p01_0 <- 332/1888 # P(Y=0,M=1|X=0)
p11_0 <- 33/1888 # P(Y=1,M=1|X=0)
p00_1 <- 1081/1918 # P(Y=0,M=0|X=1)
p10_1 <- 86/1918 # P(Y=1,M=0|X=1)
p01_1 <- 669/1918 # P(Y=0,M=1|X=1)
p11_1 <- 82/1918 # P(Y=1,M=1|X=1)
\end{Sinput}
\end{Schunk}

We proceed to compute bounds on the controlled direct effect
\(CDE(0) = P(Y(M = 0, X = 1) = 1) - P(Y(M = 0, X = 0) = 1)\) of \(X\) on
\(Y\) not passing through \(M\) at level \(M=0\), the controlled direct
effect \(CDE(1) = P(Y(M = 1, X = 1) = 1) - P(Y(M = 1, X = 0) = 1)\) at
level \(M=1\), the natural direct effect
\(NDE(0) = P(Y(M(X = 0), X = 1) = 1) - P(Y(M(X = 0), X = 0) = 1)\) of
\(X\) on \(Y\) at level \(X=0\) and the natural direct effect
\(NDE(1) = P(Y(M(X = 1), X = 1) = 1) - P(Y(M(X = 1), X = 0) = 1)\) at
level \(X=1\).

\begin{Schunk}
\begin{Sinput}
CDE0_query <- "p{Y(M = 0, X = 1) = 1} - p{Y(M = 0, X = 0) = 1}"
CDE0_obj <- analyze_graph(b, constraints = NULL, effectt = CDE0_query)
CDE0_bounds <- optimize_effect_2(CDE0_obj)
CDE0_boundsfunction <- interpret_bounds(bounds = CDE0_bounds$bounds, 
                                        parameters = CDE0_obj$parameters)
CDE0_numericbounds <- CDE0_boundsfunction(p00_0 = p00_0, p00_1 = p00_1, 
                                          p10_0 = p10_0, p10_1 = p10_1, 
                                          p01_0 = p01_0, p01_1 = p01_1, 
                                          p11_0 = p11_0, p11_1 = p11_1)

CDE1_query <- "p{Y(M = 1, X = 1) = 1} - p{Y(M = 1, X = 0) = 1}"
CDE1_obj <- update_effect(CDE0_obj, effectt = CDE1_query)
CDE1_bounds <- optimize_effect_2(CDE1_obj)
CDE1_boundsfunction <- interpret_bounds(bounds = CDE1_bounds$bounds, 
                                        parameters = CDE1_obj$parameters)
CDE1_numericbounds <- CDE1_boundsfunction(p00_0 = p00_0, p00_1 = p00_1, 
                                          p10_0 = p10_0, p10_1 = p10_1, 
                                          p01_0 = p01_0, p01_1 = p01_1, 
                                          p11_0 = p11_0, p11_1 = p11_1)
NDE0_query <- "p{Y(M(X = 0), X = 1) = 1} - p{Y(M(X = 0), X = 0) = 1}"
NDE0_obj <- update_effect(CDE0_obj, effectt = NDE0_query)
NDE0_bounds <- optimize_effect_2(NDE0_obj)
NDE0_boundsfunction <- interpret_bounds(bounds = NDE0_bounds$bounds, 
                                        parameters = NDE0_obj$parameters)
NDE0_numericbounds <- NDE0_boundsfunction(p00_0 = p00_0, p00_1 = p00_1, 
                                          p10_0 = p10_0, p10_1 = p10_1, 
                                          p01_0 = p01_0, p01_1 = p01_1, 
                                          p11_0 = p11_0, p11_1 = p11_1)

NDE1_query <- "p{Y(M(X = 1), X = 1) = 1} - p{Y(M(X = 1), X = 0) = 1}"
NDE1_obj <- update_effect(CDE0_obj, effectt = NDE1_query)
NDE1_bounds <- optimize_effect_2(NDE1_obj)
NDE1_boundsfunction <- interpret_bounds(bounds = NDE1_bounds$bounds, 
                                        parameters = NDE1_obj$parameters)
NDE1_numericbounds <- NDE1_boundsfunction(p00_0 = p00_0, p00_1 = p00_1, 
                                          p10_0 = p10_0, p10_1 = p10_1, 
                                          p01_0 = p01_0, p01_1 = p01_1, 
                                          p11_0 = p11_0, p11_1 = p11_1)
\end{Sinput}
\end{Schunk}

We obtain the same symbolic bounds as
\citep{sjolandernaturaldirecteffects} and the resulting numeric bounds
are given in Table \ref{tab:mediation-bounds} which of course agree with
those of Table V in \citep{sjolandernaturaldirecteffects}.

\begin{Schunk}
\begin{table}

\caption{\label{tab:mediation-bounds}Bounds on the controlled and natural direct effects.}
\centering
\begin{tabular}[t]{l|r|r}
\hline
  & lower & upper\\
\hline
CDE(0) & -0.20 & 0.39\\
\hline
CDE(1) & -0.78 & 0.63\\
\hline
NDE(0) & -0.07 & 0.56\\
\hline
NDE(1) & -0.55 & 0.09\\
\hline
\end{tabular}
\end{table}

\end{Schunk}

\hypertarget{a-mendelian-randomization-study-of-the-effect-of-homocysteine-on-cardiovascular-disease}{%
\subsection{A Mendelian Randomization Study of the Effect of
Homocysteine on Cardiovascular
Disease}\label{a-mendelian-randomization-study-of-the-effect-of-homocysteine-on-cardiovascular-disease}}

Mendelian randomization \citep{mendelian} assumes certain genotypes may
serve as suitable instrumental variables for investigating the causal
effect of an associated phenotype on some disease outcome.

In \citep{st0232}, the authors investigate the effect of homocysteine on
cardiovascular disease using the 677CT polymorphism (rs1801133) in the
Methylenetetrahydrofolate Reductase gene as an instrument. They use
observational data from \citep{meleady_thermolabile_2003} in which the
outcome is binary, the treatment has been made binary by a suitably
chosen cut-off at \(15\mu mol/L\), and the instrument is ternary (this
polymorphism can take three possible genotype values).

With \(X\) denoting the treatment, \(Y\) the outcome and \(Z\) the
instrument, the conditional probabilities are given as follows.

\begin{Schunk}
\begin{Sinput}
params <- list(p00_0 = 0.83, p00_1 = 0.88, p00_2 = 0.72, 
               p10_0 = 0.11, p10_1 = 0.05, p10_2 = 0.20, 
               p01_0 = 0.05, p01_1 = 0.06, p01_2 = 0.05, 
               p11_0 = 0.01, p11_1 = 0.01, p11_2 = 0.03)
\end{Sinput}
\end{Schunk}

The computation using \CRANpkg{causaloptim} is done using the following
code.

\begin{Schunk}
\begin{Sinput}
# Input causal DAG
b <- graph_from_literal(Z -+ X, Ul -+ Z, X -+ Y, Ur -+ X, Ur -+ Y)
V(b)$leftside <- c(1, 0, 1, 0, 0)
V(b)$latent <- c(0, 0, 1, 0, 1)
V(b)$nvals <- c(3, 2, 2, 2, 2)
E(b)$rlconnect <- c(0, 0,  0, 0, 0)
E(b)$edge.monotone <- c(0, 0, 0, 0, 0)
# Construct causal problem
obj <- analyze_graph(b, constraints = NULL, 
                     effectt = "p{Y(X = 1) = 1} - p{Y(X = 0) = 1}")
# Compute bounds on query
bounds <- optimize_effect_2(obj)
# Construct bounds as function of parameters
boundsfunction <- interpret_bounds(bounds = bounds$bounds, 
                                   parameters = obj$parameters)
# Insert observed conditional probabilities
numericbounds <- do.call(boundsfunction, as.list(params))
round(numericbounds, 2)
\end{Sinput}
\begin{Soutput}
#>   lower upper
#> 1 -0.09  0.74
\end{Soutput}
\end{Schunk}

Our computed bounds agree with those computed using \CRANpkg{bpbounds}
as well as those estimated using Theorem 2 of \citep{richardson2014ace},
who independently derived expressions for tight bounds that are
applicable to this setting.

\hypertarget{summary-and-discussion}{%
\section{Summary and Discussion}\label{summary-and-discussion}}

The methods and algorithms described in \citep{generalcausalbounds} to
compute symbolic expressions for bounds on non-identifiable causal
effects are implemented in the package \CRANpkg{causaloptim}. Our aim
was to provide a user-friendly interface to these methods with a
graphical interface to draw DAGs, specification of causal effects using
standard notation for potential outcomes, and an efficient
implementation of vertex enumeration to reduce computation times. These
methods are applicable to a wide variety of causal inference problems
which appear in biomedical research, economics, social sciences and
more. Aside from the graphical interface, programming with the package
is encouraged to promote reproducibility and advanced use. Our package
includes automated unit tests and also tests for correctness by
comparing the symbolic bounds derived using our program to independently
derived bounds in particular settings.

Our implementation uses a novel approach to draw DAGs using
\texttt{JavaScript} in a web browser that can then be passed to
\texttt{R} using \CRANpkg{shiny}. This graphical approach can be adapted
and used in other settings where graphs need to be specified and
computed on, such as other causal inference settings, networks, and
multi-state models. Other algorithms and data structures that could be
more broadly useful include the representation of structural equations
as \texttt{R} functions, recursive evaluation of response functions, and
parsing of string equations for causal effects and constraints.

\bibliography{causaloptim-paper.bib}

\address{%
Gustav Jonzon\\
Department of Medical Epidemiology and Biostatistics, Karolinska
Institutet\\%
\\
\url{https://ki.se/meb}\\%
\href{mailto:gustav.jonzon@ki.se}{\nolinkurl{gustav.jonzon@ki.se}}%
}

\address{%
Michael C Sachs\\
Department of Public Health, University of Copenhagen\\%
\\
\url{https://biostat.ku.dk/}\\%
\textit{ORCiD: \href{https://orcid.org/0000-0002-1279-8676}{0000-0002-1279-8676}}\\%
\href{mailto:michael.sachs@sund.ku.dk}{\nolinkurl{michael.sachs@sund.ku.dk}}%
}

\address{%
Erin E Gabriel\\
Department of Public Health, University of Copenhagen\\%
\\
\url{https://biostat.ku.dk/}\\%
\textit{ORCiD: \href{https://orcid.org/0000-0002-0504-8404}{0000-0002-0504-8404}}\\%
\href{mailto:erin.gabriel@sund.ku.dk}{\nolinkurl{erin.gabriel@sund.ku.dk}}%
}

\end{article}

\end{document}